\documentclass[pra,twocolumn,showpacs,preprintnumbers,amsmath,amssymb]{revtex4}
\usepackage{amssymb}

\usepackage{graphicx}
\usepackage{dcolumn}
\usepackage{bm}

\begin{document}


\title{Phase-imprint induced domain formations and spin dynamics in spinor condensates}

\author{Chengjun Tao}
\author{Qiang Gu}\email{qgu@sas.ustb.edu.cn}
\affiliation{Department of Physics, University of Science and
Technology Beijing, Beijing 100083, P.R. China}

\date{\today}

\begin{abstract}
We demonstrate that certain domain structures can be created both in
{\it ferro}- and {\it antiferro}-magnetic spinor condensates if the
initial phase is spatially modulated. Meanwhile, spin dynamics of
the condensate with modulated phases exhibits exotic features in
comparison with those of a condensate with a uniform phase. We
expect that these phenomena could be observed experimentally using a
phase-imprinting method.

\end{abstract}


\maketitle

Spinor Bose-Einstein condensate (BEC) has been attracting growing
attentions in the last decade since it displays a variety of exotic
phenomena associated with its spin degree of
freedom~\cite{Kurn,Ho,Stenger}. Spin-domain formation and spin
dynamics are definitely among such topics of particular interest.

Early in 1998, soon after the experimental realization of the spinor
BEC~\cite{Kurn}, the MIT group investigated the miscibility of
different spin domains in the spinor $^{23}{\rm Na}$
condensate~\cite{Stenger}. The spin-dependant interaction between
$^{23}{\rm Na}$ atoms is antiferromagnetic (AFM) and domain
structures were pre-created by applying a gradient magnetic field.
However, the $|m_F=\pm1\rangle$ domains become almost-completely
miscible as the gradient field is turned down, indicating that spin
domains is hard to be formed spontaneously in the AFM spinor
condensate. A new experimental result further rules out spontaneous
domain formation in $^{23}{\rm Na}$ condensate~\cite{Black}. Very
recently, a theoretical work demonstrates that the homogeneous
magnetic field can lead to spatial {\it modulational} instability in
AFM condensates, followed by the generation of spin
domains~\cite{Michal}. In contrast, the {\it spontaneous} domain
formation has been observed in the ferromagnetic (FM) $^{87}{\rm
Rb}$ condensate using an in-situ phase-contrast imaging~\cite{Sad}.
This is a pioneering approach in exploring spin domains in spinor
bosons. Theoretically, the domain formation is attributed to the FM
interaction between $^{87}{\rm Rb}$ atoms which leads to spontaneous
polarization in the ground state~\cite{Iso,Gu1,Zhang,Mur}.

Spin dynamics in spinor BECs, usually referring to evolutions of
spin populations in different spin
components~\cite{You,Schma1,Chang,Sch}, arises directly from spin
exchange collisions. Taking the $F=1$ manifold for example, the
collision process can be expressed as $|m_F=1\rangle +
|m_F=-1\rangle \leftrightarrow 2|m_F=0\rangle$, which naturally hold
the conservation of total spins. Owing to the quantum nature of
BECs, the collision process is coherent so that it leads to
oscillations of spin populations. Such coherent behaviors have been
observed experimentally in $^{87}{\rm Rb}$ condensates of both the
$F=1$~\cite{Chang} and $2$~\cite{Sch} manifolds, and very recently
in the AFM $^{23}{\rm Na}$ condensate~\cite{Black}. So far,
theoretical explanations for the spin dynamics are mainly based on
the assumption that each component shares the same spatial wave
function~\cite{Schma1,Chang,Sch}, called the single-mode
approximation (SMA). It seems that the SMA works well for the AFM
BEC~\cite{Black}, but becomes invalid for the FM condensate due to
its domain structures. We have ever proposed a two-domain model to
account for the later case and suggest that domain formations inside
FM BECs bring about significant influence on spin
dynamics~\cite{Gu2}.

In this paper, we propose a scheme for generating spin domains both
in FM and AFM condensates, and discuss the exotic spin dynamics
caused by domain formations. According to this scheme, domain
structures can be created by the spatially modulated phases, which
is expected to be realized via phase-imprinting method in
experiments~\cite{Dobrek}. Phase-imprinting, as a versatile tool to
manipulate BECs, has already been used to create dark
solitons~\cite{Burger}, vortices~\cite{Matt,Madison} and vortex
rings~\cite{Anderson} in scalar or two component condensates. One
can further expect that it applies to spinor-1 BECs as well.

We start with the mean-field energy functional for a spinor-1 Bose
condensate, which is expressed as~\cite{Ho}
\begin{eqnarray}\label{Eq1}
E &=& \int dr \left[  \frac{\hbar^2}{2m} \nabla\psi _i^*
\nabla\psi _i + V_{ext}(r)\psi _i^*\psi _i  \right. \nonumber\\
   &+&\frac{1}{2} c_0 \psi _i^*\psi _j^*\psi _j\psi _i
  \left. + \frac{1}{2} c_2 \psi_i^* \psi_k^*  F_{ij} F_{kl}\psi_l \psi_j
  \right]~,
\end{eqnarray}
where $\psi_\alpha$ denotes the condensate wave function for the
atomic BEC in the $\alpha$-th internal state $|m_F=\alpha\rangle$
and repeated indices are assumed to be summed. The $c_0$ and $c_2$
terms describe contributions of the spin-independent and
spin-dependent interactions between atoms respectively. The
spin-dependent interaction could be FM if $c_2<0$ or AFM if $c_2>0$.
$\textbf{\textit{F}}$ is the vector of spin matrices and
$V_{ext}(r)$ is the external trap potential. Then equations of
motion for the spinor condensate, derived from Eqn. (\ref{Eq1}) via
variational principles, are given by~\cite{Ho}
\begin{eqnarray}\label{Eq2}
{i\hbar \frac{\partial } {{\partial t}}\psi_{+1} = [{\cal H}
   + c_2 \left( {n_{+1} + n_0 - n_{-1} } \right)]\psi_{+1} + c_2 \psi_0^2 \psi_{-1}^* } ~,\nonumber\\
{i\hbar \frac{\partial } {{\partial t}}\psi _0  = [{\cal H}
   + c_2 \left( {n_{+1} + n_{-1} } \right)]\psi _0  + 2c_2 \psi_{+1}  \psi_{-1} \psi_0^* } ~,\nonumber\\
{i\hbar \frac{\partial } {{\partial t}}\psi_{-1} = [{\cal H}
   + c_2 \left( {n_{-1} + n_0 - n_{+1} } \right)]\psi_{-1} + c_2 \psi _0^2 \psi_{+1}^* }  ~,
\end{eqnarray}
where ${\cal H}=-\frac{\hbar ^2 } {2m}\nabla ^2+V_{ext}+c_0(n_{+1} +
n_0 + n_{-1})$ and $n_\alpha$ represents density of the $m_F=\alpha$
atoms. Equations (\ref{Eq2}) are just the Gross-Pitaevskii (GP)
equations for spinor-1 condensates, with the condensate wave
function expressed as $\psi_\alpha(r,t) = \sqrt{n_\alpha(r,t)}
\textit{e}^{i\theta_\alpha(r,t)}$.

The GP equations have been intensively employed to describe spin
dynamics and domain instabilities of spinor BECs, wherein the phase
$\theta$ acts as a crucial factor since it reflects the quantum
characteristics of the condensate~\cite{Zhang,Mur,You,Schma1,Chang}.
However, $\theta$ could play a more important role than recognized
previously. In previous works, $\theta$ is usually considered to be
spatially invariant. Hereinafter, we look at the case that the phase
$\theta$ can be spatially modulated. Such an extension is by no
means trivial. We will show that intriguing domain structures and
spin dynamics could be induced by appropriate phase-modulations.

\begin{figure}
\includegraphics[width=0.4\textwidth,keepaspectratio=true]{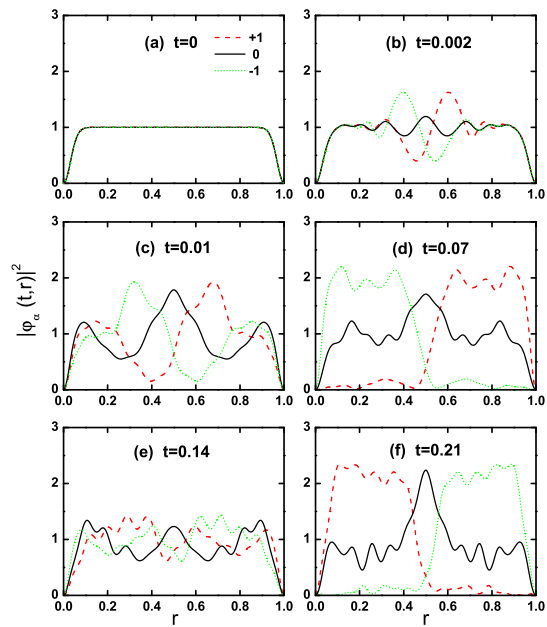}
\caption{Spin domain formation in the ferromagnetic spin-1
condensate. }\label{fig1}
\end{figure}

We consider a one-dimensional (1D) BEC as an example, supposing that
it is confined in the infinitely deep square well potential,
approximately corresponding to the elongated cigar-shaped BEC in
experiments~\cite{Gor}. The initial condensate density profile is
set to be homogeneous-like except at boundaries where it tends to
zero~\cite{Note1}, as plotted in Fig. \ref{fig1}(a). We note that
the initial phase is {\it not} homogeneous, but has been modulated
as
\begin{eqnarray}\label{Eq3}
\theta_\alpha \left( {r,0} \right) = \alpha\left( 1 - 2\theta \left(
r - L/2 \right) \right) ~,
\end{eqnarray}
where
\begin{eqnarray}\label{Eq4}
\theta \left( x \right) = \left\{ {\begin{array}{*{20}c}
   {1 \hspace{8mm} x < 0}  \\
   {0 \hspace{8mm} x > 0}  \\
\end{array} } \right.
\end{eqnarray}
is the step function and $L$ is the width of the well.

The above initial conditions are very similar to the case of
generating dark solitons in a one-dimensional scalar condensate,
where the phase modulation is realized via the phase-imprinting
method~\cite{Burger}. The scalar condensate contains only one
component, corresponding to the case of $\alpha=1$ in our model. For
spin-1 condensate, there are three individual components with
probably different initial phases, as described by Eq. (\ref{Eq3})
with $\alpha=\pm1,0$~\cite{Note2}.

For the 1D condensate, the coupling constants $c_0$ and $c_2$ can be
estimate from the 3D scattering length~\cite{Gor}. As a simple
approximation, $c_0^{1D}/c_0^{3D}\approx l/V$, where $l$ is the
elongated length of the condensate, equivalent to the width of the
well in our model, and $V$ is the condensate volume. In following
calculations, $l$ is set to be the unit of the length. And the unit
of time is defined as $t_{\rm unit} = 2ml^2/\hbar$. The effective
interaction coefficients are also re-scaled, e.g., $c_0^\prime =
2ml^2c_0/\hbar^2$ where $c_0$ is the 1D coupling constant. We note
that both $t_{\rm unit}$ and $c_0^\prime$ are proportional to the
square of the length.

\begin{figure}[b]
\includegraphics[width=0.3\textwidth,keepaspectratio=true]{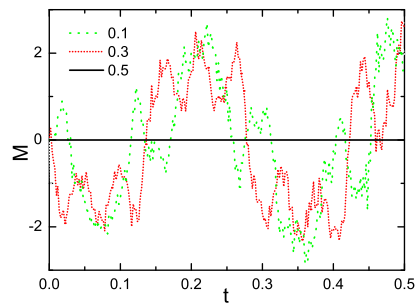}
\caption{Local magnetization density at $r=0.1$,$0.3$ and $0.5$. }
\label{fig2}
\end{figure}

First, we investigate a FM condensate, such as $^{87}$Rb. According
to the experiment data in Ref. \cite{Sad}, the $^{87}$Rb condensate
contains $2.1\times10^6$ atoms and the length of the condensate is
about $334\mu {\rm m}$. So $t_{\rm unit} \approx 300{\rm s}$ and the
re-scaled coefficients $c_0^\prime$ and $c_2^\prime$ are
approximately equal to $300$ and $-1.4$ respectively. Then dynamical
properties of the spinor condensate can be simulated using Eqs.
(\ref{Eq2}) and (\ref{Eq3}), and the results are shown in Fig.
\ref{fig1}. It can be seen that the density profiles of the three
components are changing in different manners after the evolution
begins, although they are thoroughly in the same shape initially.
The formation process of $\alpha=\pm1$ domains are apparently
demonstrated from Fig. \ref{fig1}(b) to \ref{fig1}(d). On the right
half-side of the figures, the $+1$ component overwhelms the $-1$
component, i.e., the $+1$ domain is formed. Meanwhile the $-1$
domain appears on the left. The two symmetrically-located domains
are in accord with the specific choice of initial phases as given by
Eq. (\ref{Eq3}), which implies that the domains are closely related
to the modulated phases.

Obviously, both the shape and location of domains are varying with
the time. A very interesting phenomenon occurs during the period
from Fig. \ref{fig1}(d) to (f), where the $\pm1$ domains have been
exchanging their positions. It is worth noting that the two domains
exchange their positions {\it periodically} with time. This kind of
dynamical behavior of the condensate is somewhat similar to the
soliton-like dynamics in FM spinor condensates studied by Zhang {\it
et al.} recently~\cite{Zhang2}. In Ref. \cite{Zhang2}, the
soliton-like behaviors are attributed to the exchange interaction
$c_2$, while here they are resulted from the phase modulation. As to
one domain, the inside magnetization density $M=n_{+}-n_{-}$
exhibits perfect oscillation. Figure \ref{fig2} plots the local
magnetization density $M$ at three different points in the left side
of the condensate. $M$ at $r=0.1$ and $0.3$ exhibits a good
periodicity with almost the same period within the time of our
simulation. The oscillation period is dependent on the
dimensionality of the condensate and the parameter $c_2$.
Nevertheless, at the boundary between two domains $r=0.5$, the
magnetization density is equal to $0$ and remains unchanging, owing
to the symmetry of initial density profiles and the symmetric choice
of initial phases.

\begin{figure}
\includegraphics[width=0.3\textwidth,keepaspectratio=true]{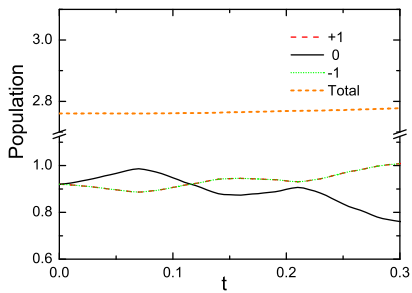}
\caption{Evolution of populations of different spin components and
total spins in a ferromagnetic condensate. } \label{fig3}
\end{figure}

Figure \ref{fig3} illustrates the evolution of total population of
each spin component (spin population) and the total spins. The spin
population smears out detailed information of inner structures
inside the condensate, but it is an important factor concerning spin
dynamics of spinor condensates and has been intensively studied
previously~\cite{Schma1,You,Chang,Sch}. It is already discovered
that the spin population exhibits a quantum oscillating feature;
this point is also confirmed in our calculations. Moreover, the
total spin conserves during the evolution, although the local
magnetization is allowed. Comparing Figs. \ref{fig2} and \ref{fig3},
one can get an interesting point that the period of population
oscillations is different from the oscillation period of the local
magnetization density. This reveals that the spin population is not
sufficient to characterize the whole feature of spin dynamics when
the condensate has certain inner structures, e.g., domain
structures.

\begin{figure}
\includegraphics[width=0.4\textwidth,keepaspectratio=true]{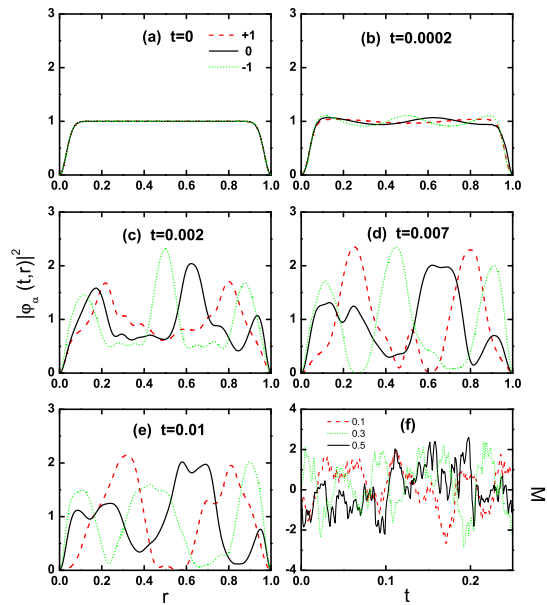}
\caption{Domain formation in a ferromagnetic spinor condensate with
phases are modulated according to Eq. (\ref{Eq5}). (a-e) show the
distribution of each component at different times. (f) shows the
evolution of the local magnetization density at different
positions.} \label{fig4}
\end{figure}

It is important to point out that domain structures can appear
spontaneously in the FM condensate even if the initial phases are
not modulated at all. This kind of domain formation is attributed to
the spontaneous symmetry breaking in the spin space cause by the FM
interaction between bosons~\cite{Iso,Gu1,Zhang,Mur} and has been
confirmed experimentally in the $^{87}{\rm Rb}$
condensate~\cite{Sad}. The spontaneous domain formation usually
leads to some multi-domain structures and domains seem distributed
randomly in the condensate~\cite{Sad,Veng}. However, the
phase-imprinting induced domain structure is strongly dependant on
the modulation of initial phases. Regular initial phases can induce
regular domain structures. As indicated above, one can even produce
a simple two-domain structure so as to facilitate experimental
probing. Moreover, if the initial phases are not that regular as
above, an irregular multi-domain structure can be produced. To
demonstrate this point, we suppose that the initial phases are
modulated in a more complicated manner, such as
\begin{eqnarray}\label{Eq5}
\theta_\alpha \left( {r,0} \right) = \sin\left((4+\alpha)\pi
r\right)
\end{eqnarray}
where different components are imprinted with different phases.
Figs. \ref{fig4}(a-e) show evolution of the condensate. A
multi-domain structure emerges and the condensate displays very
complicated dynamical behaviors. For example, the local
magnetization density shown in Fig. \ref{fig4}(f) is still
oscillating, but {\it not periodically}.

Another striking difference between the two cases is the time scale
of producing domains. The characteristic time for the spontaneous
domain formation strongly depends on the spin-dependant interaction
($t_{\rm unit} = \hbar /\left| {c_2 } \right|n$), whereas that for
phase-imprint induced domains is mainly determined by the size of
the condensate ($t_{\rm unit} = 2ml^2/\hbar$). The spontaneous
domain formation takes a period of relaxation time of nearly 100ms
in $^{87}{\rm Rb}$ condensates~\cite{Sad}. In a similar condensate
with $l=334 \mu {\rm m}$, it takes about $0.12 t_{\rm unit}\approx
36{\rm s}$ to form the two domain structure resulted from the phase
modulation, as can be seen from Fig. \ref{fig2}. In this case,
domains appear much more slowly. However, the domain formation time
can be significantly reduced by shortening the condensate length.
When the length is reduced to about $10\mu {\rm m}$, the formation
time can be decreased to within 100ms. An alternative way to reduce
the domain formation time is to enrich the phase modulation pattern.
As Fig. \ref{fig4}(f) shows, domains are well developed at $t\approx
0.01t_{\rm unit}$, much faster than the two-domain case where
$t\approx 0.12t_{\rm unit}$. In some sense, enriching the phase
modulation pattern is equivalent to reducing characteristic length
of the condensate.

\begin{figure}
\includegraphics[width=0.4\textwidth,keepaspectratio=true]{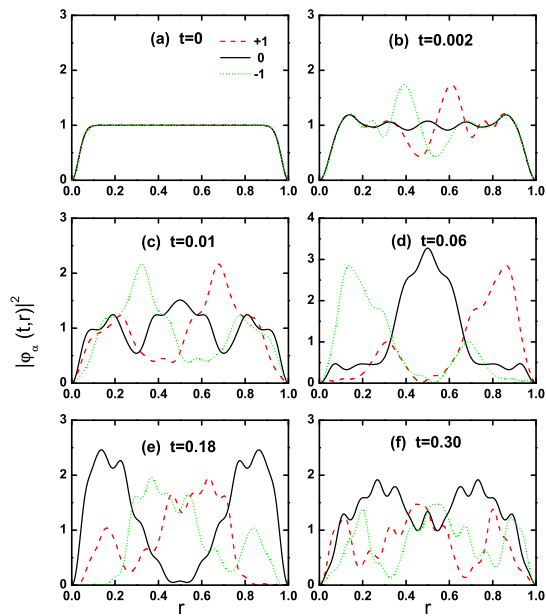}
\caption{Evolution for domain formation in a anti-ferromagnetic
condensate. Rather similar to the ferromagnetic case, the domain
structure is mainly dependent on the initial phase. } \label{fig5}
\end{figure}

In comparison with the case of FM condensate, domain formations in
AFM spinor condensates, such as $^{23}$Na, is more fascinating. A
very recent experiment suggests that the spatial domain structure
could not be formed spontaneously in the $^{23}$Na
condensate~\cite{Black}. Meanwhile, the spin dynamics measurements
show good agreement with predictions made on the base of the SMA.
Nevertheless, spin domains may be induced by applying some driving
factors. For example, a recent theory predicts that spin domains can
be generated by applying an external homogeneous magnetic field to
the AFM condensates~\cite{Michal}. So a question arises: whether
domain formation could be induced if modulating initial phases? To
answer this question, we need to simulate dynamic behaviors of the
AFM condensate according to Eqs. (\ref{Eq2}) and (\ref{Eq3}), with
the parameter $c_2$ set to be positive. We choose the parameter
$c_0^\prime$ and $c_2^\prime$ to be 100 and 10 respectively and the
obtained results are shown in Fig. \ref{fig5}. The evolution process
of AFM condensates is quite similar to the FM case, except that the
density profile of each spin component is different from that in the
FM condensate. This indicates that one can really create spin
domains in AFM condensates by the phase imprinting method, as does
in FM condensates. These simulation results await experimental
validation, for example, in the AFM $^{23}$Na condensate.

In summary, we have established that the phase-imprinting could
induce spin domains both in FM and AFM spinor condensates. Domains
come into being after the initial phases are imprinted and the
domain structure is strongly related to the spatial modulation of
phases. These characters make the phase-imprint induced domain
formation differs from the spontaneous domain formation in the FM
condensate. Even more interestingly, phase-imprinting offers the
opportunity to study spin domains and related spin dynamics in the
AFM condensate where domain structure can not form spontaneously.
Our results demonstrate that the phase engineering can play more
important roles in manipulating the quantum feature of spinor Bose
condensates than previously done.

This work is supported by the National Natural Science Foundation of
China (Grant No. 10504002), the Fok Yin-Tung Education Foundation,
China (Grant No. 101008), and the Ministry of Education of China
(NCET-05-0098).

\end{document}